\documentclass[11pt]{article}
\usepackage[preprint]{neurips_2024}

\usepackage[utf8]{inputenc}
\usepackage[T1]{fontenc}
\usepackage{hyperref}
\usepackage{url}
\usepackage{microtype}
\usepackage{xcolor}

\usepackage{CJK}
\usepackage{amsmath}
\usepackage{amssymb}
\usepackage{verbatim}
\usepackage{listings}
\usepackage{graphicx}
\usepackage{booktabs}
\usepackage{array}
\usepackage{multirow}
\usepackage{subfig}
\usepackage{pdfpages}
\usepackage{chngcntr}
\usepackage{enumitem}
\usepackage{xurl}

\setlist[enumerate,itemize]{topsep=0.5\baselineskip,itemsep=0.2\baselineskip,parsep=0pt,partopsep=0pt}
\setlist[description]{topsep=0.5\baselineskip,itemsep=0.2\baselineskip,parsep=0pt,partopsep=0pt}
\newcommand{\CHon}{\begin{CJK}{UTF8}{bkai}}
\newcommand{\CHoff}{\end{CJK}}

\renewcommand{\bibname}{References}

\newcounter{chapter}

\makeatletter
\newcommand{\neuripschapter}[1]{%
  \clearpage
  \refstepcounter{chapter}%
  \setcounter{section}{0}%
  \setcounter{subsection}{0}%
  \setcounter{subsubsection}{0}%
  \section*{#1}%
}
\newcommand{\neuripschapterstar}[1]{%
  \clearpage
  \section*{#1}%
}
\newcommand{\chapter}{\@ifstar{\neuripschapterstar}{\neuripschapter}}
\makeatother

\graphicspath{{Images/}} 
\counterwithin{figure}{chapter}

\begin{document}

%%%% Fill the basic Data here or directly edit Infos/covers.tex %%%%%%%%%%%%%%%%%%%%%%%%%%%%%%%%%%%%%%%%%%%%%%
\newcommand{\universityCH}{國立陽明交通大學}			% Normally the University name won't change
\newcommand{\universityEN}{National Yang Ming Chiao Tung University}
\newcommand{\departmentCH}{電機資訊國際學位學程}	% Type your Department Name in Chinese
\newcommand{\departmentEN}{EECS International Graduate Program}	% Type your Department Name in Chinese
\newcommand{\degreeCH}{碩士論文}				% Use this line if you are getting Master 
\newcommand{\degreeEN}{Master Thesis}
\newcommand{\titleCH}{超越單分排序：面向感知重排序實現論文推薦的可控多樣性}					% Your Thesis Title in Chinese
\newcommand{\titleEN}{Beyond Single-Score Ranking: Facet-Aware Reranking for Controllable Diversity in Paper Recommendation}			% Your Thesis Title in English
\newcommand{\nameCH}{段明濤}						% Your Chinese Name
\newcommand{\nameEN}{Aung Khant Oo}				% Your English Name
\newcommand{\AdvisorNameCH}{袁賢銘}				% Your Advisor's Chinese Name
\newcommand{\AdvisorNameEN}{Shyan-Ming Yuan} % Your Advisor's Chinese Name
\newcommand{\AdvisorNameENCover}{English, AShyan-Ming Yuan}
\newcommand{\SubmitTimeCH}{一一五年二月}			% Your graduation year/month in Chinese
\newcommand{\SubmitTimeEN}{February 2026} 

\newcommand{\SubmittedTo}{EECS International Graduate Program}% Where the Thesis was submitted to
\newcommand{\DegreeType}{Master of Science}				% Your degree type for the inside cover
\newcommand{\DegreeIn}{Electrical Engineering and Computer Science} % Your Degree's major
%%%%%%%%%%%%%%%%%%%%%%%%%%%%%%%%%%%%%%%%%%%%%%%%%%%%%%%%%%%%%%%%%%%%%%%%%%%%%%%%%%%%%%%%%%%%%%%%%%%%%%%%%%%%%%

\title{\titleEN}
\author{\nameEN \\
  \departmentEN \\
  \universityEN \\
  Advisor: \AdvisorNameEN}
\date{\SubmitTimeEN}
\maketitle

\begin{abstract}
\par
Current scientific paper recommendation systems produce a single similarity score that conflates multiple dimensions of relatedness, preventing users from specifying why papers should be similar. We present SciFACE (Scientific Faceted Cross-Encoder), a reranking framework that decomposes paper similarity into two independent dimensions: Background (what problem is addressed) and Method (how it is solved). SciFACE trains two separate cross-encoder models on 5,891 real paper pairs labeled by GPT-4o-mini with facet-specific annotations validated against human judgment. On the CSFCube benchmark, SciFACE achieves 70.63\% NDCG\%20 on Background (+5.9\% over SPECTER) and 49.06\% on Method (+31.1\% over SPECTER), matching state-of-the-art performance. Against FaBLE without citation pre-training, SciFACE outperforms by \textbf{+4.1\% NDCG\%20 on Method} using 5,891 pairs versus FaBLE's 40K synthetic augmentations, demonstrating that \textbf{high-quality grounded labels are substantially more data-efficient than large-scale synthetic augmentation} for learning fine-grained scientific similarity.

\vspace{5mm}
\end{abstract}

\linespread{1.0}\selectfont
%% it could have been called Hindu–Arabic but Arabs introduced them to Europeans
% without properly referencing their contributing sources %%%
%%%%%%% Do not forget to correctly reference the sources in your Thesis
% you see how much damage it can do ... 

%% Time to actually start writing ... you did not do any s**** so far %%
%% You should not feel you have accomplished something before at least %%
%% starting to write some text in your chapters/ChapterName.tex files %%

%% Create a ChapterName.tex source file in the folder chapters for each 
%% of your Chapter. 
%% ChapterName.tex will contain the classic Tex format style %% 
%% With for Example : 
%% \paragraph{} Intro paragraph
%% \section{First Section} Text 1
%% \subsection{Subsection} Text 2
%% \includegraphics[scale=•]{•}... \cite{ref} ... \ref{label}...etc 
%% Just as if you were using a single main.tex file, I am just splitting it
%% into different tex files because writing a thesis is a long process 

%% Personal procedure : I comment the \input{chapters/ChapterName} and type 
%% my chapter's content directly here in main.tex including the references
%% Once I am satisfied with a Chapter, I copy it in the chapters/ChapterName.tex
%% and I copy the related references in biblio/ChapterNamebib.tex 
%% Then I simply input the whole chapter with \input command
%% So I can keep a not too long main.tex and can mostly work with one file
%% The biblio/ChapterNamebib.tex is included in the Bibliography section

%%%%%%% Introduction(Considered as a Chapter for numbering)%%%%%%%%%%%%
\chapter{Introduction}
% Chapter 1: Introduction

\section{Background \& Motivation}

The volume of scientific literature has grown at an unprecedented rate. Earlier estimates report \textbf{over 2.5 million peer-reviewed articles published annually} \cite{jinha2010,ware2015}. As of 2024, arXiv reports over \textbf{670 new submissions per day} \cite{arxiv2024}, and Semantic Scholar's Academic Graph reports over \textbf{225 million papers} \cite{kinney2023}. This information overload makes it humanly impossible to comprehensively survey even a narrow sub-field. The role of intelligent recommendation systems thus shifts from a convenience to an \textbf{absolute necessity} for efficient research.

Existing paper recommendation systems typically rely on \textbf{citation networks} or \textbf{textual embeddings} to estimate relatedness. Citation-based methods such as \textbf{SPECTER} \cite{cohan2020} and \textbf{SciNCL} \cite{ostendorff2022} fine-tune pre-trained language models using triplet losses derived from citation graphs. While effective at capturing \textbf{general relatedness}, they suffer from a fundamental limitation: they produce a \textbf{single, monolithic similarity score} that conflates multiple distinct dimensions of relatedness. This single-score paradigm cannot distinguish \textbf{why} a paper is relevant, and therefore cannot serve the diverse information needs that researchers actually have:

A researcher's information need often falls into one of several distinct patterns:

\begin{enumerate}
    \item \textbf{``Same problem, different methods''} (High Background, Low Method): A researcher working on machine translation wants to compare transformer-based approaches with recurrent neural network approaches. \textit{Need:} Find papers that address the same task but use fundamentally different architectures.
    
    \item \textbf{``Same method, different problems''} (Low Background, High Method): A researcher who developed a novel graph attention mechanism wants to see how similar attention mechanisms have been applied in domains outside their expertise. \textit{Need:} Find papers that share the same core technique, regardless of application domain---enabling \textbf{cross-domain knowledge transfer}.
    
    \item \textbf{``Same problem, same methods''} (High Background, High Method): Find papers that are methodologically and topically closest as direct competitors.
\end{enumerate}

A single similarity score cannot support this level of \textbf{controllable exploration}. To address this gap, we propose decomposing paper similarity into two orthogonal dimensions:

\begin{itemize}
    \item \textbf{Background (BG):} Similarity of the research problem, task, or application domain---the ``WHAT.''
    \item \textbf{Method (MT):} Similarity of the technical approach, algorithm, or architecture---the ``HOW,'' defined in a \textbf{domain-independent} manner.
\end{itemize}

By providing separate scores for each facet, the system empowers users to filter and prioritize recommendations according to their specific exploration strategy.

\section{Task Setting}

We adopt the \textbf{Query by Example (QBE)} paradigm, where a researcher provides a seed paper and the system retrieves similar works. This approach aligns with real-world research workflows: users often start with a known relevant paper and seek to expand their bibliography by finding ``papers like this one.'' Modern academic search engines such as \textbf{Semantic Scholar} and \textbf{Google Scholar} operationalize this paradigm through their ``Related Papers'' features. In this work, we focus on \textbf{reranking} the candidate pool ($\sim$30 papers) retrieved by such QBE systems, applying cross-encoder models to produce facet-specific similarity scores.

\section{Research Question}

The primary research question driving this thesis is:

\textbf{RQ:} Can we independently model Background and Method similarity for scientific papers?

Specifically, we investigate whether a cross-encoder architecture can learn to distinguish between these facets when trained on facet-specific labels, and whether such independent modeling provides superior ranking performance compared to traditional single-score baselines.

\section{Contributions}

This thesis makes the following key contributions to the field of scientific paper recommendation:

\begin{enumerate}
    \item \textbf{Facet-Aware Reranking Framework (SciFACE):} We propose a novel reranking framework that utilizes separate cross-encoder models to score papers on Background and Method facets independently. This allows for a flexible, controllable recommendation system where users can prioritize either facet based on their current research needs. As a precision-first reranking system, SciFACE is designed to complement existing retrieval systems (e.g., Semantic Scholar API) rather than replace them, operating on small candidate pools where computational depth is most valuable.
    
    \item \textbf{Data-Efficient, High-Quality Faceted Dataset:} We construct a compact dataset of 5,891 real paper pairs across 202 seed papers, labeled by GPT-4o-mini with explicit facet-specific annotations validated against human judgment. Despite its small scale, this dataset is sufficient to train a cross-encoder model that achieves competitive performance with FaBLE---which relies on 40K synthetic pairs and a 1.3M-pair citation-pretrained base---demonstrating that \textbf{label quality can compensate for data scale} in the faceted reranking setting. This finding suggests that targeted, high-quality supervision is more informative than large-scale synthetic augmentation for learning facet-specific similarity.
    
    \item \textbf{Simple, Reproducible Methodology:} Our pipeline requires only two components---a single GPT-4o-mini annotation pass and SciBERT fine-tuning---without requiring large-scale LLM generation pipelines, citation graph preprocessing, or multi-stage hard negative mining. We further provide rigorous external validation: with \textbf{zero paper overlap} between our 5,891 training pairs and the CSFCube test set (4,207 unique papers), our results demonstrate true out-of-sample generalization. SciFACE achieves \textbf{70.63\% NDCG\%20 on Background} (+5.9\% over SPECTER) and \textbf{49.06\% on Method} (+31.1\% over SPECTER), matching state-of-the-art across both facets.
\end{enumerate}

%%%%%%%%%%%%%%%%%%%%%%%%%%%%%%%%%%%%%%%%%%%%%%%%%%%%%%%%%%%%%%%%%%%%%%%%

%%%%%%% Thesis Chapters %%%%%%%%%%%%%%%%%%%%%%%%%%%%%%%%%%%%%%%%%%%%%%%%
\chapter{Related Work}
\label{ch:related_works}
% Chapter 2: Related Work

\section{Query by Example for Academic Papers}

In this work, we focus on QBE for academic papers, specifically using \textbf{abstracts} as the query representation. Abstracts provide an ideal input for faceted similarity because they follow a relatively standardized rhetorical structure---typically containing the research \textbf{background}, the \textbf{problem} being addressed, and the \textbf{method} employed \cite{swales1990}. This implicit structure enables models to learn facet-specific representations without explicit section segmentation.

\section{Reranking in Document Retrieval}

\subsection{The Two-Stage Retrieval Pipeline}

Modern retrieval systems adopt a \textbf{two-stage architecture} to balance efficiency and precision \cite{nogueira2019}. The first stage---\textbf{retrieval}---rapidly recalls a candidate set (typically hundreds to thousands) from a large corpus using efficient methods like BM25 or approximate nearest-neighbor search. The second stage---\textbf{reranking}---applies more computationally expensive models to this smaller candidate pool, producing a refined ranking with higher precision.

This decomposition is essential because high-precision models cannot feasibly score millions of documents. Reranking thus serves as the \textbf{precision amplifier} of the pipeline, where gains in ranking quality directly translate to improved user experience.

\subsection{The Transformer Era: Pre-trained Language Models}

The introduction of \textbf{pre-trained language models} \cite{devlin2019} transformed document representation by providing \textbf{contextualized embeddings} where word meaning depends on surrounding context.

In the \textbf{bi-encoder paradigm}, query and document are independently encoded into dense vectors using a shared or separate encoder (e.g., BERT). The \texttt{[CLS]} token or pooled representation serves as the \textbf{sentence-level embedding}. Similarity is computed via dot product or cosine similarity, enabling efficient retrieval through pre-computed document embeddings and ANN search.

Prominent scientific paper embedding models include:
\begin{itemize}
    \item \textbf{SPECTER} \cite{cohan2020}: Built on SciBERT \cite{beltagy2019}, SPECTER fine-tunes the encoder using a \textbf{triplet loss} on citation graphs. Training triplets consist of (query paper, cited paper as positive, non-cited paper as negative). The model learns that citing papers are semantically related, producing document embeddings suitable for paper recommendation.
    \item \textbf{SciNCL} \cite{ostendorff2022}: Improves upon SPECTER with \textbf{neighborhood contrastive learning}, incorporating hard negatives from the citation neighborhood to learn more discriminative embeddings.
\end{itemize}

For maximum precision, \textbf{cross-encoders} concatenate query and document as a single input sequence (\texttt{[CLS] Query [SEP] Document [SEP]}) through a shared Transformer. This enables \textbf{deep token-level interaction} across the full sequence, consistently outperforming bi-encoders in ranking benchmarks \cite{nogueira2019}, albeit at a higher computational cost of $O(N)$ forward passes for $N$ candidates.

Since QBE reranking operates on $\sim$30-50 candidates from Semantic Scholar's recommendation API, cross-encoders provide optimal precision without prohibitive latency, making them our preferred architectural choice.

\section{Faceted Similarity in Scientific Literature}

A fundamental limitation of single-score similarity models is that they conflate multiple distinct dimensions of relatedness into a single scalar. A paper may be similar to another in terms of the \textbf{problem} it addresses, the \textbf{method} it employs, or the \textbf{results} it achieves---yet these dimensions are conceptually orthogonal. Recent research has begun to address this multi-faceted nature of scientific similarity.

\subsection{CSFCube: The First Faceted QBE Benchmark}

Mysore et al. introduced \textbf{CSFCube} (Computer Science Faceted Query by Example Cube) \cite{mysore2021}, the first expert-annotated benchmark for faceted scientific document retrieval. The dataset comprises 50 diverse query papers drawn from computational linguistics and machine learning venues, each paired with a candidate pool assembled via depth-$k$ pooling ($k \in \{100, 250\}$) from multiple retrieval systems including TF-IDF, SPECTER, and citation-based methods. Expert annotators assigned graded relevance scores (0--3) for each query-candidate pair independently across three facets: \textbf{Background} (same research problem or domain), \textbf{Method} (same technical approach or algorithm), and \textbf{Result} (similar empirical findings).

Notably, SPECTER achieves only $\sim$37\% NDCG\%20 on the Method facet---far below its Background performance---revealing that citation-based training fails to encode mechanistic similarity. More broadly, papers frequently receive high scores on one facet and low on another, confirming that facets are orthogonal dimensions of scientific similarity and exposing a fundamental ceiling on any single-score model.

\subsection{ASPIRE: Multi-Vector Representations}

Mysore et al. proposed \textbf{ASPIRE} (Aspect-based Scientific Paper Information Retrieval) \cite{mysore2022} as a direct response to the single-score limitation of SPECTER. ASPIRE adds multiple projection heads on top of a shared SciBERT encoder, each trained with a facet-specific contrastive loss derived from structured citation contexts---specifically, whether a citation appears in the background, method, or result section of the citing paper. Two variants, TSASPIRE (optimal transport alignment) and OTASPIRE (multi-match training), further refine correspondence between facet embeddings. Despite these advances, ASPIRE remains fundamentally a bi-encoder: facet embeddings are computed independently for each document, limiting the model's ability to capture fine-grained pairwise interactions. Moreover, its training signal still originates from citation structure, which is an imperfect proxy for true faceted relevance.

\subsection{FaBLE: LLM-Based Facet Blending}

Do et al. introduced \textbf{FaBLE} (Facet Blending) \cite{do2024} to address the data scarcity problem in faceted paper retrieval. Rather than relying on citation signals, FaBLE uses LLaMA-2 to segment paper abstracts into facet-specific chunks (Background, Method, Result), then generates synthetic training pairs by recombining segments across papers---pairs with swapped Method segments, for instance, receive a high Background but low Method label. This augmentation strategy produces large-scale supervised data without human annotation, and the best FaBLE variant (with hard negatives and SPECTER initialisation) achieves $\sim$49\% NDCG\%20 on the CSFCube Method facet. However, the synthetic nature of these pairs means the training distribution may not reflect the complexity of real pairwise comparisons, and the pipeline requires a capable LLM for segmentation, adding significant preprocessing overhead.

\subsection{Limitations of Prior Work}

Despite meaningful progress, the above approaches share two structural limitations. First, all models---SPECTER, SciNCL, ASPIRE, and FaBLE---adopt a bi-encoder architecture that encodes each document independently, precluding the token-level cross-document attention that is essential for fine-grained similarity judgement. Second, their training signals, whether derived from citation graphs or LLM-generated synthetic pairs, are indirect approximations of true faceted relevance rather than direct human-validated supervision. SciFACE addresses both limitations: by adopting a cross-encoder architecture and training on compact but high-quality GPT-4o-mini annotations validated against human judgement, it enables richer pairwise interaction modelling with a more faithful supervision signal.

We focus on Background and Method facets as the primary axes of controllable literature exploration---``what problem?'' versus ``how solved?''---and exclude the Result facet, whose relevance is tied to specific experimental outcomes that are often incompletely or inconsistently described in abstracts.
%%%%%%%%%%%%%%%%%%%%%%%%%%%%%%%%%%%%%%%%%%%%%%%%%%%%%%%%%%%%%%%%%%%%%%%%

\chapter{Methodology}
\label{ch:methodology}
% Chapter 3: Methodology

\begin{figure}[!ht]
\centering
\includegraphics[width=\textwidth]{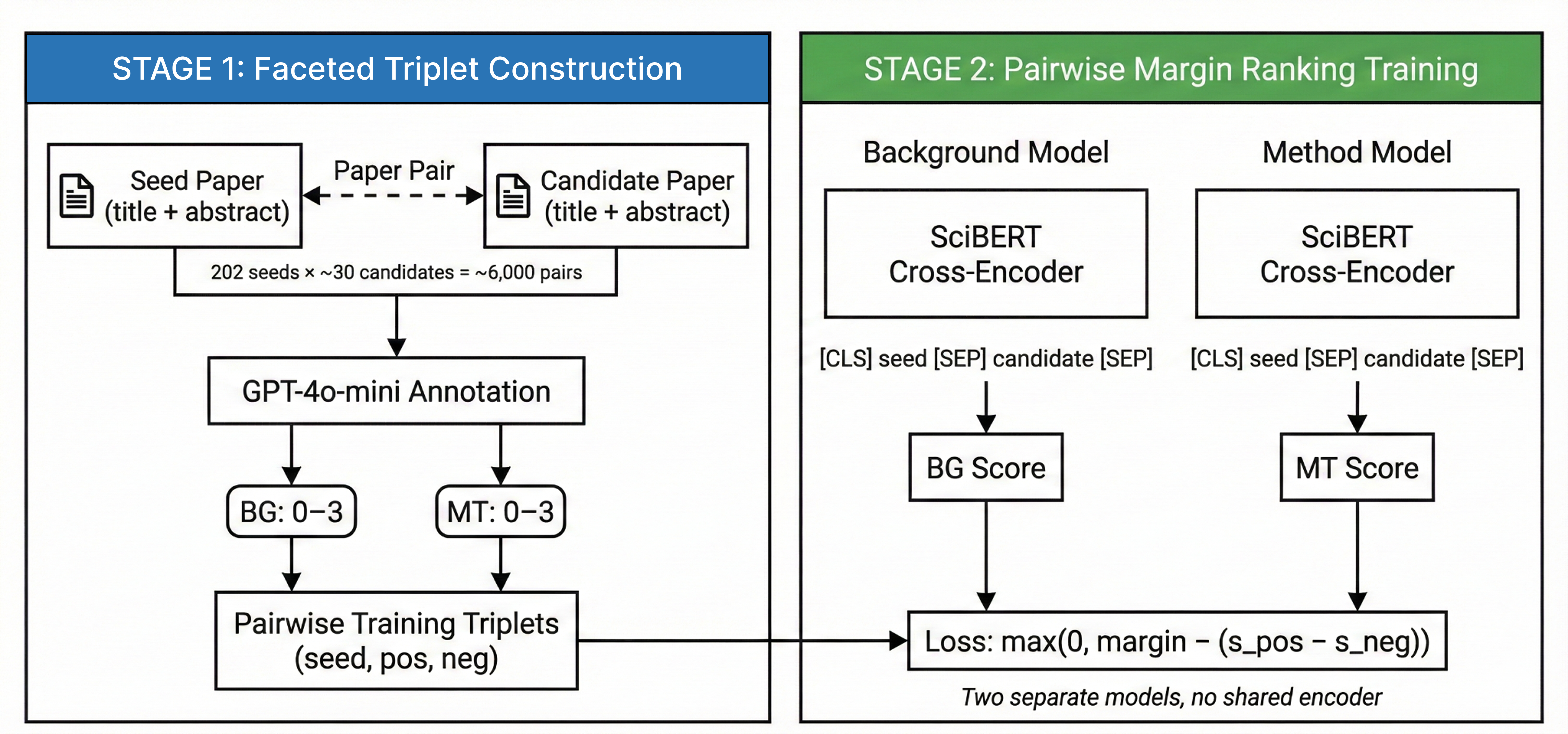}
\caption{System overview of our facet-aware reranking pipeline. Stage 1 constructs facet-labeled triplets from seed-candidate pairs using GPT-4o-mini (Background and Method scores). Stage 2 trains two independent SciBERT cross-encoders (one for Background, one for Method) using pairwise margin ranking.}
\label{fig:system_overview}
\end{figure}

This chapter presents our facet-aware reranking methodology, structured around the two-stage pipeline depicted in Figure~\ref{fig:system_overview}. We begin by formulating the paper recommendation task and motivating our pairwise ranking approach (\S~\ref{sec:problem_formulation}). The chapter then follows the primary stages of our system: \textbf{Stage 1: Faceted Triplet Construction} (\S~\ref{sec:stage1}), which encompasses data collection, facet definitions, LLM annotation, and pair generation; and \textbf{Stage 2: Pairwise Margin Ranking Training} (\S~\ref{sec:stage2}), which details the facet-specific cross-encoder architectures and the optimization process.

\section{Problem Formulation \& Design Rationale}
\label{sec:problem_formulation}

We formulate the paper recommendation task as a \textbf{facet-aware reranking} problem. Given a user-provided seed paper $q$ and a candidate paper $c$ from a retrieved pool $\mathcal{C}$ (where $|\mathcal{C}| \approx 30$), the goal is to predict two independent similarity scores: $S_{BG}(q, c)$ and $S_{MT}(q, c)$.

Faceted similarity judgments are inherently subjective. Even trained human annotators on the CSFCube benchmark achieve only moderate agreement (Spearman $\rho$=0.63 after expert adjudication) \cite{mysore2021}. This subjectivity is intrinsic to the task, making absolute score prediction (pointwise regression) unreliable due to varying annotation standards. Therefore, we adopt a \textbf{pairwise ranking} objective. Rather than predicting exact scores, we train the model to correctly \textit{order} candidates by relative similarity. Pairwise comparison is more robust because it only requires ordinal agreement (e.g., that candidate $A$ is more similar to $q$ than candidate $B$), avoiding the noise of absolute calibration.

To train with this pairwise objective, our system must dynamically construct $(q, c^+, c^-)$ triplets where $c^+$ is more similar to the query than $c^-$. The following sections detail the two main stages of our pipeline required to generate these triplets and train the ranking models.

%======================================================================
\section{Stage 1: Faceted Triplet Construction}
\label{sec:stage1}
%======================================================================

The first stage of our pipeline aims to construct high-quality training triplets by retrieving seed-candidate pairs and scoring them on independent facets using an LLM.

\subsection{Seed Selection \& Candidate Retrieval}
\label{sec:data_collection}

Learning domain-independent method similarity requires seeds where the same method appears across different research areas. We constructed the seed set through two strategies: keyword-based search on arXiv using method-oriented queries (e.g., ``graph neural network,'' ``vision transformer''), and citation-based expansion via the Semantic Scholar API. Seeds were filtered to include only papers whose abstracts clearly describe both a research problem and a technical approach. This yielded 202 seed papers across five domain categories (Table~\ref{tab:seed_domains}). Graph Neural Networks form the largest group because GNN-based methods naturally appear across molecular science, social networks, and vision---providing rich cross-domain method pairs.

\begin{table}[h]
\centering
\caption{Distribution of seed papers by domain.}
\label{tab:seed_domains}
\begin{tabular}{lr}
\toprule
Domain & Seeds \\
\midrule
Graph Neural Networks & 56 \\
Computer Vision & 44 \\
NLP / Text & 42 \\
Reinforcement Learning & 22 \\
Other (Robotics, Medical AI, etc.) & 38 \\
\midrule
Total & 202 \\
\bottomrule
\end{tabular}
\end{table}

For each seed, we retrieved 30 candidate papers via the Semantic Scholar Recommendation API\footnote{\url{https://api.semanticscholar.org/api-docs/recommendations}}, which returns semantically related papers not necessarily cited by the seed. After removing duplicates and candidates with missing abstracts, this yields \textbf{5,891 seed-candidate pairs}. Having collected these pairs, the next step is to score each one by facet---so that we know which candidate should serve as $c^+$ and which as $c^-$ in a training triplet. We obtain these scores through LLM annotation, described below.

\subsection{LLM Annotation}
\label{sec:llm_labeling}

We score each pair on two independently defined facets, grounded in the annotation philosophy of CSFCube~\cite{mysore2021}. \textbf{Background (BG)}---``What is this paper about?''---captures similarity in research problem, task, and application domain. \textbf{Method (MT)}---``How does this paper solve the problem?''---captures similarity in technical approach, \emph{independent of application domain}. Table~\ref{tab:facet_criteria} operationalizes each facet on a 0--3 scale; edge rules apply (MT=0 if no clear method is described; MT$\leq$1 if papers share only a high-level paradigm such as ``deep learning'').

\begin{table}[!ht]
\centering
\caption{Facet Labeling Criteria}
\label{tab:facet_criteria}
\begin{tabular}{cll}
\toprule
Score & Background (WHAT problem?) & Method (HOW solved?) \\
\midrule
3 & Same exact task/objective & Same core architecture (domain-independent) \\
2 & Related tasks in same sub-area & Same method family, different architecture \\
1 & Same ML problem type, diff domain & Specific shared mechanism \\
0 & Unrelated communities & Different approaches OR not stated \\
\bottomrule
\end{tabular}
\end{table}

Each pair was labeled by GPT-4o-mini (\texttt{temperature=0}) using a structured prompt (Appendix~\ref{app:prompt}). The model receives the titles and abstracts of both papers and outputs two independent scores (0--3) for Background and Method, along with a one-sentence justification. The prompt enforces three constraints: (1) explicit domain-independent method scoring with calibration examples, to prevent the model from conflating domain distance with method distance; (2) conservative scoring (``If unsure, choose the lower score''); and (3) structured JSON output requiring reasoning.

To validate quality, we conducted iterative independent manual annotation on a 100-pair sample. Early rounds revealed an LLM bias toward inflating Method scores for same-domain pairs. Refinements to the prompt's domain-independence calibration examples successfully resolved this. The final Spearman's $\rho$ agreement between human and GPT-4o-mini labels is reported in Table~\ref{tab:annotation_agreement}, alongside CSFCube's reference baseline.

\begin{table}[!ht]
\centering
\caption{Annotation Agreement (Spearman $\rho$)}
\label{tab:annotation_agreement}
\begin{tabular}{lcc}
\toprule
Comparison & Background & Method \\
\midrule
Human vs. GPT-4o-mini (Ours) & 0.61 & 0.44 \\
CSFCube Pre-Adjudication & 0.45 & 0.31 \\
\bottomrule
\end{tabular}
\end{table}

\subsection{Triplet Generation}
\label{sec:pair_construction}

With each seed-candidate pair now carrying a BG score and an MT score, we construct training triplets by exploiting the relative ordering of candidates. For a given seed $q$ and its annotated candidates $\{c_1, \ldots, c_n\}$, we enumerate all pairs $(c_i, c_j)$ where $\text{label}(c_i) > \text{label}(c_j)$, independently for each facet: $c_i$ becomes $c^+$ and $c_j$ becomes $c^-$. Because BG and MT scores are assigned independently, the same pair $(c_i, c_j)$ can contribute to both triplet sets with reversed roles---for instance, if $c_i$ scores higher on BG but lower on MT than $c_j$, it acts as $c^+$ in a BG triplet and $c^-$ in an MT triplet. The dataset is split at the seed level (Train: 4,136 / Val: 900 / Test: 855) to prevent data leakage.

One preprocessing refinement applies to the Method facet: we collapse the 4-point scale (0--3) into a 3-point scale (0--2) by merging scores 2 and 3. As shown in Table~\ref{tab:score_dist}, score 3 constitutes only 4.7\% of MT-labeled pairs, making the distinction between ``exact same architecture'' (score 3) and ``same method family'' (score 2) unreliable from abstract text alone. Merging these two levels improved downstream ranking performance; a detailed ablation is provided in Section~\ref{sec:mt_ablation}.

\begin{table}[!ht]
\centering
\caption{Score Distribution Across All Labeled Pairs (n=5,891)}
\label{tab:score_dist}
\begin{tabular}{ccccc}
\toprule
Score & BG Count & BG \% & MT Count & MT \% \\
\midrule
0 & 408 & 6.9 & 1,075 & 18.2 \\
1 & 2,136 & 36.3 & 1,837 & 31.2 \\
2 & 1,954 & 33.2 & 2,705 & 45.9 \\
3 & 1,393 & 23.6 & 274 & 4.7 \\
\bottomrule
\end{tabular}
\end{table}

Notably, the dataset captures the cross-domain variation central to our task: \textbf{16.3\%} of pairs are ``Cross-Domain Same-Method'' (BG~$\leq 1$, MT~$\geq 2$), and \textbf{22.6\%} are ``Same-Domain Different-Method'' (BG~$\geq 2$, MT~$\leq 1$). These patterns constitute hard negatives that single-score models cannot distinguish, forming the contrastive backbone of our training data.

%======================================================================
\section{Stage 2: Pairwise Margin Ranking Training}
\label{sec:stage2}
%======================================================================

In the second stage, we leverage the generated triplets to train two specialized ranking models.

\subsection{Facet-Specific Cross-Encoders}
\label{sec:model_architecture}

We employ SciBERT \cite{beltagy2019} (110M parameters) as our base encoder due to its domain-specific vocabulary tailored for scientific literature. Unlike bi-encoders that encode documents independently, our cross-encoder takes the concatenated seed-candidate pair as a single input:
\begin{center}
\texttt{[CLS] Seed Title [SEP] Seed Abstract [SEP] Candidate Title [SEP] Candidate Abstract [SEP]}
\end{center}
The sequence is truncated at 512 tokens. Truncation occurs at the abstract tail to preserve titles and opening sentences, which typically front-load the most critical methodological information. The \texttt{[CLS]} representation passes through dropout ($p=0.1$) and a linear layer to output a scalar similarity score.

Following established paradigms (e.g., ASPIRE), we explicitly train two independent models ($M_{BG}$ and $M_{MT}$) rather than a multi-head architecture. Background similarity aligns well with topical overlap, whereas Method similarity demands detecting structural analogies across semantic gulfs. These distinct objectives yield conflicting training signals. Our preliminary experiments with multi-head training showed degraded performance, confirming that each facet requires dedicated representational capacity.

\subsection{Model Optimization}
\label{sec:training}

We train the model with MarginRankingLoss:
\begin{equation}
\mathcal{L} = \max(0, \text{margin} - (S(q, c^+) - S(q, c^-)))
\end{equation}
This pushes the model to score the positive candidate ($c^+$) consistently higher than the negative ($c^-$) by at least a margin of 0.5. To balance training efficiency with exposure to diverse comparisons, training pairs are resampled at the start of each epoch, sub-sampling up to 10 $(c^+, c^-)$ triplets per seed. 

Optimization utilizes AdamW (learning rate $2\times 10^{-5}$) with a linear warmup over 10\% of the steps, followed by linear decay. We employ gradient clipping (max norm 1.0) and a batch size of 16 over 10 epochs. The best checkpoints are selected via validation set Spearman correlation.

%%%%%%%%%%%%%%%%%%%%%%%%%%%%%%%%%%%%%%%%%%%%%%%%%%%%%%%%%%%%%%%%%%%%%%%%

\chapter{Experiments}
\label{ch:experiments}
% Chapter 4: Experiments

\section{Evaluation Setup}

We evaluate our models on the \textbf{CSFCube} benchmark \cite{mysore2021}, a gold-standard dataset for faceted query-by-example.
\begin{itemize}
    \item \textbf{Benchmark Characteristics:} CSFCube contains 50 queries (16 BG, 17 MT, 17 Result) with human-annotated relevance judgments (0-3 scale).
    \item \textbf{External Validation:} Crucially, there is \textbf{zero paper overlap} between our training data (5,891 pairs) and the CSFCube benchmark (4,207 unique papers), ensuring that our results represent true out-of-sample generalization.
    \item \textbf{Baselines:} We compare against \textbf{SPECTER} \cite{cohan2020}, a strong bi-encoder baseline trained on citation graphs.
\end{itemize}
Following the standard CSFCube evaluation protocol, we report \textbf{NDCG\%20} (primary metric) and \textbf{MAP}.

\section{Quantitative Results}

We present a comprehensive comparison against multiple baselines on the CSFCube benchmark. Baseline results for all FaBLE and ASPIRE variants are taken from FaBLE~\cite{do2024} under identical evaluation settings.

\begin{table}[!ht]
\centering
\caption{SciFACE vs. Baselines on CSFCube (Background \& Method facets). All baseline results are reproduced from FaBLE. $\dagger$ denotes models initialized from SPECTER-COCITE$_{\text{Spec}}$, a bi-encoder pre-trained on 1.3M co-citation pairs. SciFACE uses only 5,891 GPT-labeled real pairs with no citation-based pre-fine-tuning.}
\label{tab:main_results}
\begin{tabular}{lcccc}
\toprule
\multirow{2}{*}{Model} & \multicolumn{2}{c}{Background} & \multicolumn{2}{c}{Method} \\
\cmidrule(lr){2-3} \cmidrule(lr){4-5}
 & NDCG\%20 & MAP & NDCG\%20 & MAP \\
\midrule
SPECTER~\cite{cohan2020}                        & 66.70 & 43.95 & 37.41 & 22.44 \\
+FaBle~\cite{do2024}                            & 67.38 & 42.66 & 44.97 & 25.98 \\
TSASPIRE$^\dagger$~\cite{mysore2022}            & 70.22 & 49.58 & 48.20 & 28.86 \\
OTASPIRE$^\dagger$~\cite{mysore2022}            & \textbf{71.04} & 50.56 & 46.46 & 27.64 \\
SPECTER-COCITE$_{\text{Spec}}$                  & 70.03 & 49.99 & 45.99 & 25.60 \\
+FaBle\_Spec$^\dagger$~\cite{do2024}            & 70.09 & 45.93 & 49.14 & 30.90 \\
+FaBle\_Spec +HN$^\dagger$~\cite{do2024}        & 69.48 & 46.03 & \textbf{49.43} & \textbf{32.57} \\
\midrule
\textbf{SciFACE (Ours)}                 & 70.63 & \textbf{52.16} & \textbf{49.06} & 26.40 \\
\bottomrule
\end{tabular}
\end{table}

SciFACE achieves competitive performance on Background (70.63\% NDCG\%20, +5.9\% over SPECTER, highest MAP at 52.16\%) and matches state-of-the-art on Method (49.06\%, +31.1\% over SPECTER), using only \textbf{5,891 real labeled pairs}---no citation graphs, no synthetic augmentation, no citation-pretrained base model.

The most striking finding concerns data efficiency. SciFACE approaches the best FaBLE variant (+FaBle\_Spec +HN, 49.43\%) on the Method facet while requiring approximately \textbf{200$\times$ fewer actual training pairs} (5,891 vs.\ 1.3M+). Compared against FaBLE on a fairer footing---using the same standard SPECTER base without citation pre-training---SciFACE outperforms by \textbf{+4.1\% NDCG\%20 on Method} (49.06\% vs.\ 44.97\%). We attribute this gap to label quality: FaBLE synthesizes training data via LLaMA-2 text generation, which produces plausible but potentially noisy pseudo-documents, while SciFACE trains on real peer-reviewed abstracts with human-validated GPT-4o-mini labels (Spearman $\rho$=0.44 for Method, exceeding CSFCube's pre-adjudication baseline of 0.31). These results support the hypothesis that \textbf{label quality compensates for data scale} in the faceted reranking setting.

One metric requires clarification. SciFACE's MAP on Method (26.40) is lower than FaBLE+HN (32.57). This is expected: MAP applies a binary relevance threshold, whereas our pairwise ranking objective optimizes graded ordering. NDCG\%20, which natively respects graded relevance, is the primary metric for CSFCube and the appropriate basis for comparison.

\section{Qualitative Analysis}

This section analyzes systematic failure patterns to understand the model's limitations and boundaries.

\subsection{Error Analysis}
\label{sec:error_analysis}

To understand the model's limitations, we conducted a systematic analysis of 82 false negative cases---paper pairs with high ground truth relevance (GT $\geq$ 2) that the model ranked outside the top 10. Two recurring error patterns emerged.

\textbf{Pattern 1: Granularity Mismatch.} The model fails to recognize methodological similarity when papers share high-level algorithmic concepts but differ in concrete implementations. Shared theoretical vocabulary (``graph,'' ``structure,'' ``optimization'') is insufficient for the model to bridge the gap between different instantiations of the same algorithmic family.

\begin{table}[!ht]
\centering
\caption{Granularity Mismatch Example}
\label{tab:granularity_mismatch}
\begin{tabular}{p{11cm}cc}
\toprule
Query / Candidate Pair & GT & Our Rank \\
\midrule
\textbf{Query:} Edge-Linear First-Order Dependency Parsing with Undirected MST Inference \newline
\textbf{Candidate:} Finding Optimal Bayesian Network Given a Super-Structure & 2 & \#87/250 \\
\bottomrule
\end{tabular}
\end{table}

Both papers address structure learning via graph optimization, but the query proposes an efficient MST algorithm for NLP dependency parsing, while the candidate develops a constrained search for Bayesian network learning. Despite belonging to the same algorithmic family (graph-based structure optimization), the model fails to detect their methodological kinship.

\textbf{Pattern 2: Cross-Domain Terminology Barrier.} The model fails to match methodologically equivalent approaches when they are described using entirely different domain-specific vocabulary.

\begin{table}[!ht]
\centering
\caption{Cross-Domain Terminology Barrier Example}
\label{tab:terminology_barrier}
\begin{tabular}{p{11cm}cc}
\toprule
Query / Candidate Pair & GT & Our Rank \\
\midrule
\textbf{Query:} Compact Personalized Models for Neural Machine Translation \newline
\textbf{Candidate:} Direct adaptation of hybrid DNN/HMM model for fast speaker adaptation & 2 & \#92/96 \\
\bottomrule
\end{tabular}
\end{table}

Both papers propose compact, parameter-efficient adaptation techniques---the query uses group lasso regularization for NMT, while the candidate uses speaker codes for speech recognition. The core methodology (freezing most parameters, adapting via small offsets) is equivalent, but the domain-specific framing creates near-zero lexical overlap.

\subsection{Summary}

The qualitative analysis demonstrates both the model's strengths and its boundaries:
\begin{itemize}
    \item \textbf{Strengths:} The model successfully decouples ``what is studied'' from ``how it is studied,'' achieving cross-domain method matching when sufficient lexical overlap exists between abstracts.
    \item \textbf{Limitations:} Performance degrades when (1) high-level algorithmic terms mask implementation differences, or (2) equivalent methods are described in domain-specific vocabulary with minimal lexical overlap.
\end{itemize}

\section{Ablation Study: Method Score Granularity}
\label{sec:mt_ablation}

We investigate the impact of label granularity on Method facet performance. As described in Section~\ref{sec:pair_construction}, we merge the original 4-point scale (0--3) into a 3-point scale (0--2) by collapsing scores 2 and 3.

\begin{table}[!ht]
\centering
\caption{Method Facet: Label Scale Ablation (CSFCube Test Set)}
\label{tab:mt_ablation}
\begin{tabular}{lc}
\toprule
Scale & NDCG\%20 \\
\midrule
0--3 (original) & 45.57 \\
0--2 (merged) & \textbf{49.06} \\
\bottomrule
\end{tabular}
\end{table}

\textbf{Analysis:} Merging scores 2 and 3 improves NDCG\%20 by \textbf{+3.5 percentage points} (45.6\% $\rightarrow$ 49.1\%). This confirms our hypothesis: the distinction between ``exact same architecture'' (score 3) and ``same method family'' (score 2) is too subtle to reliably learn from abstract text alone, and the sparse score-3 class (only 4.7\% of labels) introduces noise. The coarser 3-point scale provides cleaner training signal.

%%%%%%%%%%%%%%%%%%%%%%%%%%%%%%%%%%%%%%%%%%%%%%%%%%%%%%%%%%%%%%%%%%%%%%%%

\chapter{Discussion}
\label{ch:discussion}
% Chapter 5: Discussion

\section{Findings}

The results of this study validate our core hypothesis: \textbf{scientific paper similarity is not monolithic.} By explicitly decoupling ``Background'' and ``Method,'' we achieved state-of-the-art ranking performance---using a fraction of the data required by competing approaches.

\begin{itemize}
    \item \textbf{The ``Method Gap'' Reduced:} The most significant finding is the substantial improvement (+31.1\% NDCG\%20) on the Method facet. This confirms that citation-based models fail to capture procedural similarity, a gap SciFACE closes via cross-encoder token-level interaction and targeted facet supervision.
    
    \item \textbf{Data Efficiency of High-Quality Labels:} SciFACE is competitive with the highest-performing FaBLE variant (+FaBle\_Spec +HN) on Method using \textbf{only 5,891 real labeled pairs}, compared to 1.3M co-citation pairs plus 40K synthetic augmentations. When compared against +FaBle (40K synthetic pairs, no citation-pretrained base), SciFACE outperforms by +4.1\% NDCG\%20. This confirms that \textbf{targeted, validated LLM annotations on real abstracts provide a stronger training signal} than large-scale synthetic generation.
    
    \item \textbf{Validity of LLM Labeling:} The strong generalization from GPT-4o-mini labels (internal training) to the CSFCube benchmark (human expert annotations) confirms that modern LLMs can serve as reliable annotators for complex scientific concepts, given careful prompt engineering and validation against human judgment.
\end{itemize}

\section{Design Trade-offs}

SciFACE makes deliberate design choices that frame its advantages and scope of use:

\begin{itemize}
    \item \textbf{Reranking over Retrieval (Precision-first):} Cross-encoders require $O(N)$ pairwise inference, making them suited for reranking small candidate pools ($\sim$30 papers) rather than large-scale retrieval. This is an intentional design decision: by operating as a precision amplifier at the final reranking stage, SciFACE avoids the approximation costs inherent to bi-encoder architectures. The computational constraint does not limit applicability, as modern academic search APIs (e.g., Semantic Scholar) already perform first-stage retrieval efficiently.

    \item \textbf{Simple, Reproducible Pipeline:} SciFACE's pipeline consists of two steps---GPT-4o-mini annotation and SciBERT fine-tuning---with no multi-stage preprocessing, specialized segmentation models, or citation-pretrained base required. This simplicity lowers the barrier to reproduction and adaptation to new scientific domains.
    
    \item \textbf{Lexical Dependency:} As shown in our error analysis (\S~\ref{sec:error_analysis}), the model struggles when methodologically similar papers use entirely different terminology (cross-domain barrier) or when high-level algorithmic terms mask implementation differences (granularity mismatch). This is a known limitation of abstract-level encoding.
    
    \item \textbf{Facet Coverage:} We model Background and Method---the primary axes of controllable literature exploration. Other facets such as ``Results'' (performance claims) or ``Data'' (datasets used) are left for future work.
\end{itemize}

\section{Future Work}

\begin{itemize}
    \item \textbf{Bi-Encoder Distillation:} Knowledge distillation from SciFACE's high-precision Cross-Encoder into a Bi-Encoder or Multi-Vector architecture (e.g., ColBERT) would enable efficient large-scale retrieval while preserving facet-aware ranking quality.
    \item \textbf{Method Taxonomy Injection:} Incorporating explicit method taxonomies or LLM-extracted canonical method descriptors (e.g., mapping ``group lasso regularization'' and ``speaker code adaptation'' to the same ``parameter-efficient adaptation'' family) could bridge the cross-domain terminology barrier.
    \item \textbf{Hierarchical Method Representation:} A hierarchical encoding scheme that represents both algorithmic families (e.g., ``graph-based optimization'') and task-specific instantiations (e.g., ``MST parsing'', ``Bayesian network search'') separately could improve recall for granularity mismatch cases.
    \item \textbf{Extending to ``Result'' Facet:} Extending the framework to include a Result facet would allow users to find papers with similar empirical findings, completing the Background--Method--Result triad established by CSFCube.
\end{itemize}

%%%%%%%%%%%%%%%%%%%%%%%%%%%%%%%%%%%%%%%%%%%%%%%%%%%%%%%%%%%%%%%%%%%%%%%%

%%%%%%%% Conclusion(Considered as a Chapter for numbering)%%%%%%%%%%%%%%
\chapter{Conclusion}
\label{ch:conclusion}
% Chapter 6: Conclusion

This thesis presented \textbf{SciFACE}, a facet-aware reranking framework for scientific paper recommendation. By decomposing similarity into \textbf{Background} and \textbf{Method} dimensions and training specialized Cross-Encoders on a novel LLM-labeled dataset, we demonstrated that it is possible to support controllable, fine-grained literature exploration. 

Our models achieved substantial improvements over the SPECTER baseline on the expert-annotated CSFCube benchmark---particularly on the Method facet (+31.1\% NDCG\%20)---while matching state-of-the-art performance using only 5,891 GPT-labeled pairs validated against human judgment on a sampled subset. This work represents a step towards the next generation of academic search engines---ones that understand not just \textit{that} two papers are related, but exactly \textit{how} and \textit{why}.

%%%%%%%%%%%%%%%%%%%%%%%%%%%%%%%%%%%%%%%%%%%%%%%%%%%%%%%%%%%%%%%%%%%%%%%%

%%%%%%%% Bibliography(Excluded from numbering) %%%%%%%%%%%%%%%%%%%%%%%%%
%% Bibliography uses 1.5cm linespace %%%%
\linespread{1.3}\selectfont

% %% On NYCU Library Website Page about Thesis Format, each Department is %%
%% bounded to a citation style like ACM or IEEE, you should check yours %%
%% EECS use IEEE citation style %%
\clearpage
\renewcommand{\bibname}{References}
\addcontentsline{toc}{chapter}{References}
\bibliographystyle{plainnat}
\bibliography{Bibliographies/egbib}
%%%% Insert the Appendices if any %%%%%%%%%%%%%%%%%%%%%%%%
\appendix
\chapter{Appendix}

\section{Labeling Prompt}
\label{app:prompt}
The complete system prompt used for GPT-4o-mini annotation is shown below:

\begin{small}
\begin{verbatim}
You will score similarity between a Seed paper and a Candidate paper 
on TWO INDEPENDENT facets (0-3):
- Background = similarity of research problem/task (WHAT)
- Method = similarity of technical approach (HOW), DOMAIN-INDEPENDENT

CRITICAL:
- Method is DOMAIN-INDEPENDENT. Same technique across different domains 
  can still be Method=3.
- Do NOT mix facets: task/goal -> Background; 
  technique/architecture/training paradigm -> Method.
- Be conservative: if unsure, choose the LOWER score.

---
BACKGROUND (0-3) = WHAT problem/task is solved?
- 3: Same exact task/objective 
     (e.g., NER vs NER; node classification vs node classification)
- 2: Closely related tasks in same sub-area 
     (e.g., node classification vs link prediction; MT vs summarization)
- 1: Same ML problem type but different application domains/communities 
     (e.g., NLP vs CV; molecules vs social networks)
- 0: Separated research communities / unrelated problems

METHOD (0-3) = HOW is it solved? (IGNORE domain!)
- 3: Same core architecture/technique, including direct variants or 
     improvements that keep the same fundamental approach 
     (e.g., GCN, MBGCN, GCN+attention all count as "GCN"; 
      BERT, RoBERTa, ALBERT all count as "BERT")
- 2: Same method family but different core architectures 
     (e.g., GCN vs GraphSAGE; BERT vs GPT; ResNet vs VGG)
- 1: ONLY if you can NAME a specific shared core mechanism 
     (NOT generic "deep learning"/"attention"/"embedding"). 
     Examples: contrastive learning, message passing, diffusion models, 
     policy gradient RL, meta-learning
- 0: Fundamentally different approaches OR method not clearly stated

EDGE RULES:
- If abstract does NOT clearly state method/technique -> Method=0
- If multiple methods are mentioned -> score the PRIMARY method only

---
OUTPUT (JSON only, no extra text):
{
  "background": {"score": 0-3, "reason": "Compare problems/tasks."},
  "method": {"score": 0-3, "reason": "Compare techniques, ignoring domain."}
}

---
CRITICAL EXAMPLES:

Example 1 [Cross-domain, same method]:
Seed: "Graph Convolutional Networks for molecular property prediction"
Candidate: "Graph Convolutional Networks for social network analysis"
-> Background: 1 (both prediction, different domains)
-> Method: 3 (both use GCN)

Example 2 [Same task, no shared method]:
Seed: "BERT for named entity recognition"
Candidate: "CRF-based named entity recognition"
-> Background: 3 (same task: NER)
-> Method: 0 (BERT vs CRF, no shared mechanism)

Example 3 [Method variants = Method 3]:
Seed: "Graph Convolutional Networks for node classification"
Candidate: "Multi-view Block-wise GCN for node classification"
-> Background: 3 (same task)
-> Method: 3 (MBGCN is a direct variant of GCN)

Example 4 [Method family = Method 2]:
Seed: "Graph Convolutional Networks for node classification"
Candidate: "GraphSAGE for node classification"
-> Background: 3 (same task)
-> Method: 2 (GCN vs GraphSAGE are same family, different architectures)

---
Reminder:
- Variant/improvement of same architecture -> Method = 3
- Same family but different architecture -> Method = 2
- Same technique + different domain -> Method can be 3; Background often 1
\end{verbatim}
\end{small}

\section{Seed Selection Keywords}
\label{app:keywords}
The following method-oriented keywords were used for arXiv search during seed selection:

\begin{itemize}
    \item Graph Neural Networks: ``graph neural network'', ``graph convolutional network'', ``message passing neural network''
    \item Computer Vision: ``vision transformer'', ``convolutional neural network'', ``object detection''
    \item NLP: ``transformer'', ``language model'', ``named entity recognition'', ``machine translation''
    \item Reinforcement Learning: ``policy gradient'', ``Q-learning'', ``actor-critic''
    \item General: ``contrastive learning'', ``self-supervised learning'', ``diffusion model''
\end{itemize}

%----------------------------------------------------------------------
\section{Annotation Quality Validation}
\label{app:annotation_validation}

We validated GPT-4o-mini annotations by independently re-annotating 100 randomly sampled pairs using human judgment. This section presents detailed agreement analysis.

\subsection{Agreement Comparison}

Table~\ref{tab:agreement_comparison} compares our human--LLM agreement with CSFCube's pre-adjudication inter-annotator agreement.

\begin{table}[!ht]
\centering
\caption{Spearman $\rho$ Agreement Comparison}
\label{tab:agreement_comparison}
\begin{tabular}{lcc}
\toprule
Setting & Background & Method \\
\midrule
Human vs. GPT-4o-mini (Ours) & 0.61 & 0.44 \\
CSFCube Pre-Adjudication & 0.45 & 0.31 \\
\bottomrule
\end{tabular}
\end{table}

Our annotations achieve higher agreement than CSFCube's pre-adjudication baseline, particularly for the Background facet.

\subsection{Score Distribution}

Table~\ref{tab:score_distribution} shows the distribution of scores assigned by human annotators versus GPT-4o-mini.

\begin{table}[!ht]
\centering
\caption{Score Distribution Comparison (n=100)}
\label{tab:score_distribution}
\begin{tabular}{ccccc}
\toprule
\multirow{2}{*}{Score} & \multicolumn{2}{c}{Background} & \multicolumn{2}{c}{Method} \\
\cmidrule(lr){2-3} \cmidrule(lr){4-5}
 & Human & GPT & Human & GPT \\
\midrule
0 & 32 & 5 & 67 & 20 \\
1 & 31 & 51 & 20 & 19 \\
2 & 24 & 23 & 5 & 58 \\
3 & 13 & 21 & 8 & 3 \\
\bottomrule
\end{tabular}
\end{table}

Key observations:
\begin{itemize}
    \item \textbf{Background:} Humans assign more 0s (32 vs 5), indicating stricter standards for unrelated problems.
    \item \textbf{Method:} GPT-4o-mini over-assigns score 2 (58 vs 5), while humans are more conservative, assigning score 0 to 67\% of pairs.
\end{itemize}

\subsection{Confusion Matrices}

\begin{table}[!ht]
\centering
\caption{Background Confusion Matrix (Human $\times$ GPT-4o-mini)}
\label{tab:bg_confusion}
\begin{tabular}{c|cccc}
\toprule
Human $\backslash$ GPT & 0 & 1 & 2 & 3 \\
\midrule
0 & \textbf{5} & 24 & 3 & 0 \\
1 & 0 & \textbf{18} & 9 & 4 \\
2 & 0 & 5 & \textbf{11} & 8 \\
3 & 0 & 4 & 0 & \textbf{9} \\
\bottomrule
\end{tabular}
\end{table}

\begin{table}[!ht]
\centering
\caption{Method Confusion Matrix (Human $\times$ GPT-4o-mini)}
\label{tab:mt_confusion}
\begin{tabular}{c|cccc}
\toprule
Human $\backslash$ GPT & 0 & 1 & 2 & 3 \\
\midrule
0 & \textbf{20} & 15 & 32 & 0 \\
1 & 0 & \textbf{2} & 18 & 0 \\
2 & 0 & 0 & \textbf{5} & 0 \\
3 & 0 & 2 & 3 & \textbf{3} \\
\bottomrule
\end{tabular}
\end{table}

The Method confusion matrix reveals that GPT-4o-mini frequently assigns score 2 when humans assign 0 (32 cases), suggesting the model identifies superficial method similarities that human annotators do not consider meaningful.

\subsection{Disagreement Analysis}

We identified 38 pairs with $|\text{diff}| \geq 2$ on at least one facet. Common disagreement patterns include:

\begin{enumerate}
    \item \textbf{Method granularity mismatch:} GPT-4o-mini assigns MT=2 for papers sharing generic mechanisms (e.g., ``graph neural networks''), while humans require more specific architectural similarity.
    
    \item \textbf{Domain vs. method conflation:} Despite prompt calibration, GPT-4o-mini occasionally conflates domain similarity with method similarity for same-domain papers.
    
    \item \textbf{Abstract ambiguity:} When abstracts lack explicit method descriptions, GPT-4o-mini infers methods from context, while humans default to MT=0.
\end{enumerate}
%%%%%%%%%%%%%%%%%%%%%%%%%%%%%%%%%%%%%%%%%%%%%%%%%%%

%%%% Insert the Autobiography (not necessary) %%%%%%
% \chapter*{Autobiography}
% \addcontentsline{toc}{chapter}{Autobiography}
% \input{chapters/autobiography}
%%%%%%%%%%%%%%%%%%%%%%%%%%%%%%%%%%%%%%%%%%%%%%%%%%%
\end{document}